# Low latency global carbon budget reveals strong land sink recovery in 2025

Philippe Ciais[1,*], Piyu Ke[1,*], Xiangjun Tian[2,*], Stephen Sitch[3], Wei Li[4], Xiaomeng Du[4], Xiaofan Gui[5], Ben Poulter[6], Thomas Colligan[6], Auke M. van der Woude[7], Anne-Wil van den Berg[7], Wouter Peters[7], Zhu Liu[4], Zhu Deng[8], Zhe Jin[9], Yilong Wang[2], Junjie Liu[10], Sudhanshu Pandey[10], Chris O'Dell[11], Jiang Bian[5], John Miller[12], Xin Lan[12,13], Jefferson Goncalves De Souza[3], Michael O'Sullivan[3], Pierre Friedlingstein[3,14], Youngryel Ryu[15], Helin Zhang[15], Julien Alléon[1], Yi Xi[1], Daniel S. Goll[1], Lei Zhu[1], Guido R. van der Werf[7], Yitong Yao[16], Shilong Piao[9], Frédéric Chevallier[1]

1. Laboratoire des Sciences du Climat et de l'Environnement, University Paris Saclay CEA CNRS, Gif sur Yvette 91191, France
2. State Key Laboratory of Tibetan Plateau Earth System, Environment and Resources (TPESER), Institute of Tibetan Plateau Research, Chinese Academy of Sciences, Beijing 100101, China
3. Faculty of Environment, Science and Economy, University of Exeter, Exeter EX4 4QF, United Kingdom
4. Department of Earth System Science, Ministry of Education Key Laboratory for Earth System Modeling, Institute for Global Change Studies, Tsinghua University, Beijing 100084, China
5. Machine learning group, Microsoft research, Beijing 100080, China
6. Spark Climate Solutions, Covina, CA, USA
7. Environmental Sciences Group, Dept of Meteorology and Air Quality, Wageningen University, Wageningen 6708 PB, the Netherlands
8. Department of Geography, The University of Hong Kong, Hong Kong SAR, China
9. Institute of Carbon Neutrality, Sino-French Institute for Earth System Science, College of Urban and Environmental Sciences, Peking University, Beijing 100871, China
10. Jet Propulsion Laboratory, California Institute of Technology, Pasadena 91011, CA, USA
11. Cooperative Institute for Research in the Atmosphere, Colorado State University, Fort Collins, CO 80523, USA
12. National Oceanic and Atmospheric Administration Global Monitoring Laboratory, CO 80303, USA
13. Cooperative Institute for Research in Environmental Sciences, University of Colorado Boulder, CO 80303, USA
14. Laboratoire de Météorologie Dynamique, IPSL, CNRS, ENS, Université PSL, Sorbonne Université, École Polytechnique, Paris 75005, France
15. Department of Landscape Architecture and Rural Systems Engineering, College of Agriculture and Life Sciences, Seoul National University, Seoul 08826, Republic of Korea
16. Institute of Environment and Ecology, Tsinghua Shenzhen International Graduate School, Tsinghua University, Shenzhen 518055, China

***Correspondence:** philippe.ciais@cea.fr (P.C.), kepiyu@gmail.com (P.K.), tianxj@itpcas.ac.cn (X.T.)

In 2025, the atmospheric $CO_2$ growth rate based on the globally averaged marine boundary layer (MBL) observations from the NOAA network fell to 2.06 ± 0.09 ppm $yr^{-1}$ after the record-high 3.76 ± 0.09 ppm $yr^{-1}$ in 2024 and below the 2015–2022 mean of 2.47 ppm $yr^{-1}$ (Fig. 1a) [1]. Both the satellite-based Growth Rates Using Satellite Observations (GRESO) estimates, derived from independent OCO-2 and GOSAT observations [2], and our OCO-2-constrained flux inversions showed a marked decline in the whole-atmosphere $CO_2$ growth rate between 2024 and 2025, from 3.40 ± 0.09 to 1.96 ± 0.09 ppm $yr^{-1}$ and from 3.63 ± 0.05 to 2.15 ± 0.05 ppm $yr^{-1}$, respectively. The monthly GRESO record shows that this decline developed mainly from late 2024 into early 2025, following the decay of the 2023/24 El Niño and coinciding with a moderation of the exceptional positive global temperature anomaly, with monthly $CO_2$ growth remaining comparatively low through much of 2025 (Fig. 1b). The MBL $CO_2$ growth rate declined by 45.2% between 2024 and 2025, even as global fossil $CO_2$ emissions were estimated to increase by 0.7% to 10.38 GtC $yr^{-1}$ in 2025 [3,4]. This contrasting evolution points to a recovery of terrestrial carbon uptake after the pronounced land sink weakening documented in 2024 [3,5]. Here, we present a low-latency global and regional carbon budget for the year 2025, using top-down inversions and bottom-up models, with a new satellite-driven biophysical model BESS [6]. We assess the contributions of land and ocean carbon uptake to the sharp decline in atmospheric $CO_2$ growth and quantify the magnitude and spatial pattern of land sink recovery following the 2023–2024 weakening.

We combined bottom-up land flux estimates from three dynamic global vegetation models (DGVMs), JULES, ORCHIDEE-MICT and LPJ-EOSIM [7–9], with machine-learning emulators of the ocean biogeochemical models and data-driven air-sea products used in the Global Carbon Budget (GCB) 2025 [3,10], and top-down land and ocean flux estimates from four atmospheric inversion models (CAMS, CMS-Flux, CTE and GONGGA), using OCO-2 retrievals of total column $CO_2$ [11–14], which improves data coverage, particularly over the tropics, with a data release latency of about six weeks, providing better spatial resolution and earlier detection of flux anomalies than in situ networks. A riverine carbon transport correction was applied consistently to both land and ocean fluxes from all four atmospheric inversions. The DGVMs typically simulate fires with prognostic fire modules using population density, lightning ignitions, climate and simulated fuel moisture but have weaknesses in capturing extreme forest fires and tropical forest degradation and deforestation fires. Hence, for the two DGVMs that simulated fire (ORCHIDEE-MICT and LPJ-EOSIM), we subtracted their original fire emissions and then added the mean of the Global Fire Emissions Database (GFED4.1s) [15] and the Global Fire Assimilation System (GFAS) [16] for the period 2010-2025 to all three DGVMs. Our approach allows a rapid diagnosis of the most recent changes in the global carbon cycle although we use the ERA5 climate forcing for DGVMs in this study instead of CRU-JRA in the GCB protocol, which has a time latency of around one year. Net land uptake is defined here as the sum of non-fossil land $CO_2$ fluxes including photosynthesis, respiration, fire and land-use change emissions.

# The global carbon budget in 2025

Fig. 1c shows the bottom-up carbon budget, obtained by combining the average fossil emissions estimated by Carbon Monitor and previous projections of the Global Carbon Budget 2025 [3,4] with our bottom-up estimates of net land and ocean carbon fluxes. Fig. 1d shows the top-down carbon budget based on the mean of the four OCO-2 inversions. Global fossil $CO_2$ emissions were 10.38 GtC $yr^{-1}$ with a range of 10.34-10.42 GtC $yr^{-1}$ from our two estimates. In the bottom-up budget, the DGVMs estimated a net land sink of 2.04 ± 0.24 GtC $yr^{-1}$ in 2025, a marked reversal from the net land source of 0.98 ± 0.51 GtC $yr^{-1}$ in 2024. In the top-down budget, the four atmospheric inversions estimated a net land sink of 2.68 ± 0.20 GtC $yr^{-1}$ in 2025, compared with only a small sink of 0.08 ± 0.51 GtC $yr^{-1}$ in 2024. Averaging the bottom-up and top-down estimates gives a global net land sink of 2.36 ± 0.16 GtC $yr^{-1}$ in 2025. This represents a strengthening of 2.81 ± 0.31 GtC $yr^{-1}$ from 2024, with the 2025 sink 0.71 ± 0.13 GtC $yr^{-1}$ stronger than the 2015–2022 mean. The ocean $CO_2$ uptake for 2025 was 3.11 ± 0.36 GtC $yr^{-1}$ (our ocean flux emulators: 3.15 ± 0.65 GtC $yr^{-1}$, inversions: 3.06 ± 0.32 GtC $yr^{-1}$), similar to the 2024 mean. The bottom-up budget imbalance, defined as fossil emissions minus the sinks estimated by bottom-up models and the observed $CO_2$ growth rate, is 0.81 ± 0.69 GtC $yr^{-1}$ using the growth rate of MBL stations and a conversion factor of 2.124 GtC per ppm. As expected, the top-down budget closed to within 0.07 ± 0.26 GtC $yr^{-1}$ using the atmospheric accumulation implied by the inversions. The contrasting changes in land and ocean uptake indicate that the rebound of the terrestrial sink is the dominant carbon budget change associated with the sharp decline in atmospheric $CO_2$ growth in 2025.

# Regional carbon flux anomalies in 2025

To gain insights into which regions contributed to the recovery of the land sink in 2025, we analyzed quarterly spatial patterns of land and ocean $CO_2$ flux anomalies from the bottom-up models and OCO-2 inversions, using 2015–2022 as the reference period. At the global scale, ocean uptake remained close to the 2015–2022 mean, in contrast to the much larger change in the land sink. The quarterly land and ocean flux anomalies are shown in Fig. S1, and their relationship with terrestrial water-storage anomalies is displayed in Fig. S2.

Over the ocean, both the emulator ensemble and the inversions indicated slightly stronger global carbon uptake relative to the 2015–2022 mean, with sink anomalies of 0.26 ± 0.22 and 0.06 ± 0.31 GtC $yr^{-1}$, respectively (Fig. S1). The clearest common regional signal was stronger uptake in the Southern Ocean, defined following the RECCAP2 ocean regional mask, with anomalies of 0.11 ± 0.10 GtC $yr^{-1}$ in the emulators and 0.12 ± 0.16 GtC $yr^{-1}$ in the inversions. The largest regional difference occurred in the Pacific Ocean, where the emulator ensemble indicated a small sink anomaly of 0.04 ± 0.09 GtC $yr^{-1}$, whereas the OCO-2 inversions indicated a source anomaly of 0.09 ± 0.17 GtC $yr^{-1}$ relative to the reference mean.

Over the land, the regional net $CO_2$ flux anomalies in Fig. S1 showed a marked recovery of tropical carbon uptake in 2025. At the ensemble-mean level, tropical lands shifted from net $CO_2$ sources in 2024 to net sinks in 2025, from 1.70 ± 0.28 to 0.85 ± 1.23 GtC $yr^{-1}$ in the DGVMs and from 1.35 ± 0.79 to 0.48 ± 0.67 GtC $yr^{-1}$ in the inversions. Relative to the 2015–2022 mean, the 2025 tropical land flux was a stronger sink by 0.32 ± 1.19 GtC $yr^{-1}$ in the DGVMs and 1.08 ± 0.44 GtC $yr^{-1}$ in the inversions, although the larger spread among DGVMs indicates greater uncertainty in the magnitude of the modelled tropical rebound. Both approaches showed enhanced uptake across large parts of Africa and northern Eurasia, while larger differences remained over Latin America, East Asia and North America (Fig. S1). In the Northern Hemisphere extratropics (>23.5°N), the DGVMs retained a source anomaly of 0.19 ± 1.21 GtC $yr^{-1}$ in 2025 relative to 2015-2022 whereas the inversions indicated a sink anomaly of 0.28 ± 0.38 GtC $yr^{-1}$.

Over the tropical lands, the quarterly anomalies indicate a different timing of recovery between the two approaches (Fig. S1). The DGVMs produced weak source anomalies during the first half of 2025 and shifted toward stronger than normal uptake during the second half of the year, whereas the inversions indicated enhanced tropical uptake throughout 2025. Over tropical South America, source anomalies estimated by the inversions persisted in parts of the region during JFM 2025 following the severe drought and fire-related carbon losses of late 2024 [5], but weakened thereafter. The inversions showed a broader shift toward sink anomalies from AMJ 2025 onward, whereas the DGVMs retained a more spatially heterogeneous mixture of source and sink anomalies (Fig. S1). In contrast, Africa showed widespread regions of enhanced uptake in both approaches. The spatial distribution of the 2025 land flux anomalies also covaried with terrestrial water storage from GRACE, with wetter conditions frequently coinciding with enhanced $CO_2$ uptake and drier conditions with reduced uptake in both the DGVMs and inversions (Fig. S2).

To determine whether the 2025 sink rebound occurred where the land sink had previously weakened, we quantified recovery in grid cells where the mean 2023–2024 sink loss relative to 2015–2022 exceeded 10 gC $m^{-2}$ land $yr^{-1}$ (Fig. 1e,f). Such a pattern could reflect short-term memory in the land carbon cycle, for example through the refilling of fast pools such as litter and foliage that were depleted in 2023–2024. Recovery of 100% denotes a return to the 2015–2022 mean flux, whereas values above 100% indicate that land uptake in 2025 exceeded the reference mean. Within areas that experienced substantial carbon sink losses in 2023-2024, about 80% of the land area showed some recovery in 2025 in both approaches (80.1% in the DGVMs and 80.0% in the inversions). Overall recovery reached 87.3% in the DGVMs and 99.5% in the inversions. Recovery was strongest in the tropics, where it exceeded 100% in both approaches (DGVMs: 106.4%; inversions: 116.4%), but remained weaker in the Northern Hemisphere extratropics (DGVMs: 59.6%; inversions: 90.7%). Nearly half of the substantial-loss area showed recovery exceeding 100% (48.5% in the DGVMs and 48.9% in the inversions). Thus, the global rebound reflected incomplete recovery in some regions but land uptake exceeding the 2015–2022 mean in others, rather than a spatially uniform recovery.

Together, these results show that the sharp decline in atmospheric $CO_2$ growth rate in 2025 was driven primarily by a strong rebound in terrestrial carbon uptake, but that this recovery was spatially uneven. Tropical recovery generally restored or exceeded the reference sink strength, whereas recovery remained incomplete across northern extratropical regions. Due to the recovery being primarily in storage in fast turnover pools, i.e., litter and foliage, the variability in atmospheric $CO_2$ growth is likely to remain with terrestrial systems as the primary driver. The strong global land sink in 2025 therefore reflected a broad recovery, with uptake exceeding the 2015–2022 mean globally despite incomplete recovery in parts of the northern extratropics.

# Data availability

The data from Global Carbon Budget 2025 are available at https://globalcarbonbudget.org/datahub/the-latest-gcb-data-2025/. The OCO-2 retrievals are available at disc.gsfc.nasa.gov/datasets?page=1&keywords=OCO-2. NOAA/GML $CO_2$ data are available at https://gml.noaa.gov/ccgg/trends/. The Carbon Monitor fossil emissions dataset is available at carbonmonitor.org. The GFED 4.1s fire emissions dataset is available at geo.vu.nl/~gwerf/GFED/GFED4/. The GFAS fire emissions dataset is available at atmosphere.copernicus.eu/global-fire-monitoring/. The ERA5 monthly averaged data is available at cds.climate.copernicus.eu/cdsapp#!/dataset/reanalysis-era5-single-levels-monthly-means?tab=overview. The GRACE/FO TWS data used in this study are available at www2.csr.utexas.edu/grace/RL06_mascons.html.

# Author contributions

P.C. and P.K. designed the research; P.K. and P.C. performed the analysis; P.K. collected and analyzed the research data; P.C. and P.K. created the first draft of the paper; all authors contributed to the interpretation of the results and to the text.

**Conflict of interest statement.** None declared.

# Acknowledgements

We acknowledge the Global Carbon Project, which is responsible for the Global Carbon Budget. We thank the ocean modelling and f$CO_2$-mapping groups for producing and making available their model and f$CO_2$-product output, and the land modelling groups for producing and making available their model output. We also thank the ICOS Carbon Portal (ICOS ERIC) for providing access to the dataset 'Global $CO_2$ gridded flux fields from 14 atmospheric inversions in GCB2025' (DOI: 10.18160/VZ3B-GHYQ). P.C., F.C., S.S., P.F. acknowledge support from the European Space Agency Climate Space RECCAP2-CS project (ESA ESRIN/4000144908), ESA Carbon-RO (4000140982/23/I-EF) and the CALIPSO and CLARiTy projects. Support for the CALIPSO and CLARiTy projects was provided by Schmidt Sciences, LLC. P.C. received funding from the European Union's Horizon Europe research and innovation programme under grant agreement No 101081395 (EYE-CLIMA). D.S.G. received support through the CLARiTy project. Work of J.L. and S.P. was conducted at the Jet Propulsion Laboratory, California Institute of Technology, under a contract with the National Aeronautics and Space Administration (80NM0018D0004). Additionally, J.L. acknowledges the funding support from NASA Orbiting Carbon Observatory Science Team program. The CAMS simulations were granted access to the HPC resources of TGCC under the allocation A0190102201 made by GENCI. This work was also supported by a computing grant from the Dutch national e-infrastructure with the support of the SURF Cooperative (NWO-2025.010). This research was supported in part by the NOAA cooperative agreement NA22OAR4320151. The statements, findings, conclusions, and recommendations are those of the author(s) and do not necessarily reflect the views of NOAA or the U.S. Department of Commerce. Y.R. was supported by National Research Foundation of Korea (RS-2024-00348585) and Ministry of Climate, Energy and Environment (202300218237).

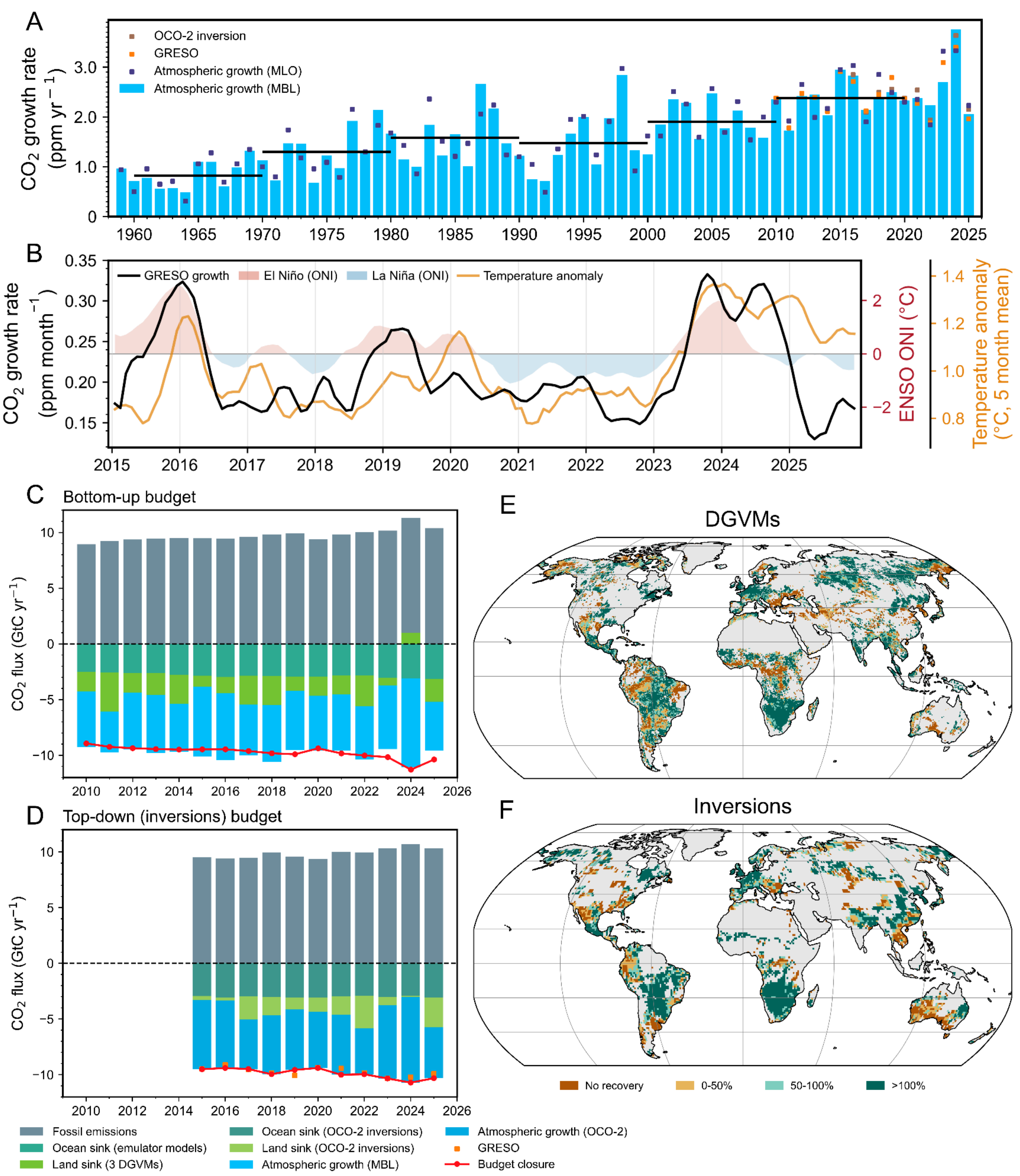

**Fig. 1 Atmospheric $CO_2$ growth rate from 1959-2025, global carbon budget from 2010-2025 and spatial recovery of the 2023-2024 land sink losses in 2025.** (a) Annual atmospheric $CO_2$ growth rates from globally averaged marine boundary layer observations (MBL, blue bars), the Mauna Loa station (MLO, dark blue squares), OCO-2 constrained atmospheric inversions (brown squares), and the Growth Rates Using Satellite Observations approach (GRESO, orange squares). GRESO values are based on GOSAT for 2010–2014 and on the mean of independently derived OCO-2 and GOSAT estimates from 2015 onward. (b) Monthly GRESO $CO_2$ growth rate (black line), Oceanic Niño Index (ONI; red and blue shading for El Niño and La Niña conditions, respectively), and global surface temperature anomaly (orange line, 5-month running mean) from 2015 to 2025. (c) Global bottom-up

$CO_2$ budget obtained from fossil $CO_2$ emissions, net land $CO_2$ flux from three DGVMs, ocean $CO_2$ uptake from the low latency ocean emulator ensemble, and atmospheric $CO_2$ accumulation derived from MBL observations. The red curve is -1 * fossil emissions and the difference between the bars and this curve is the imbalance of the bottom-up budget. (d) Global top-down $CO_2$ budget obtained from fossil $CO_2$ emissions and four atmospheric inversions constrained by OCO-2 observations. Atmospheric accumulation is derived from the inversion ensemble; orange squares show the corresponding budget sum using the independent GRESO atmospheric growth rate estimate. Spatial recovery in 2025 of the mean 2023–2024 net land $CO_2$ sink loss relative to the 2015–2022 baseline, estimated from the three DGVMs (e) and four atmospheric inversions (f). Recovery is evaluated only where the mean 2023–2024 sink loss exceeds 10 gC $m^{-2}$ land $yr^{-1}$. Brown indicates no recovery, yellow 0–50% recovery, light green 50–100% recovery, and dark green >100% recovery. A recovery of 100% denotes a return to the 2015–2022 mean flux, whereas values >100% indicate that the 2025 land sink exceeded the reference mean. Grey land areas did not meet the substantial loss threshold or were excluded from the recovery analysis.

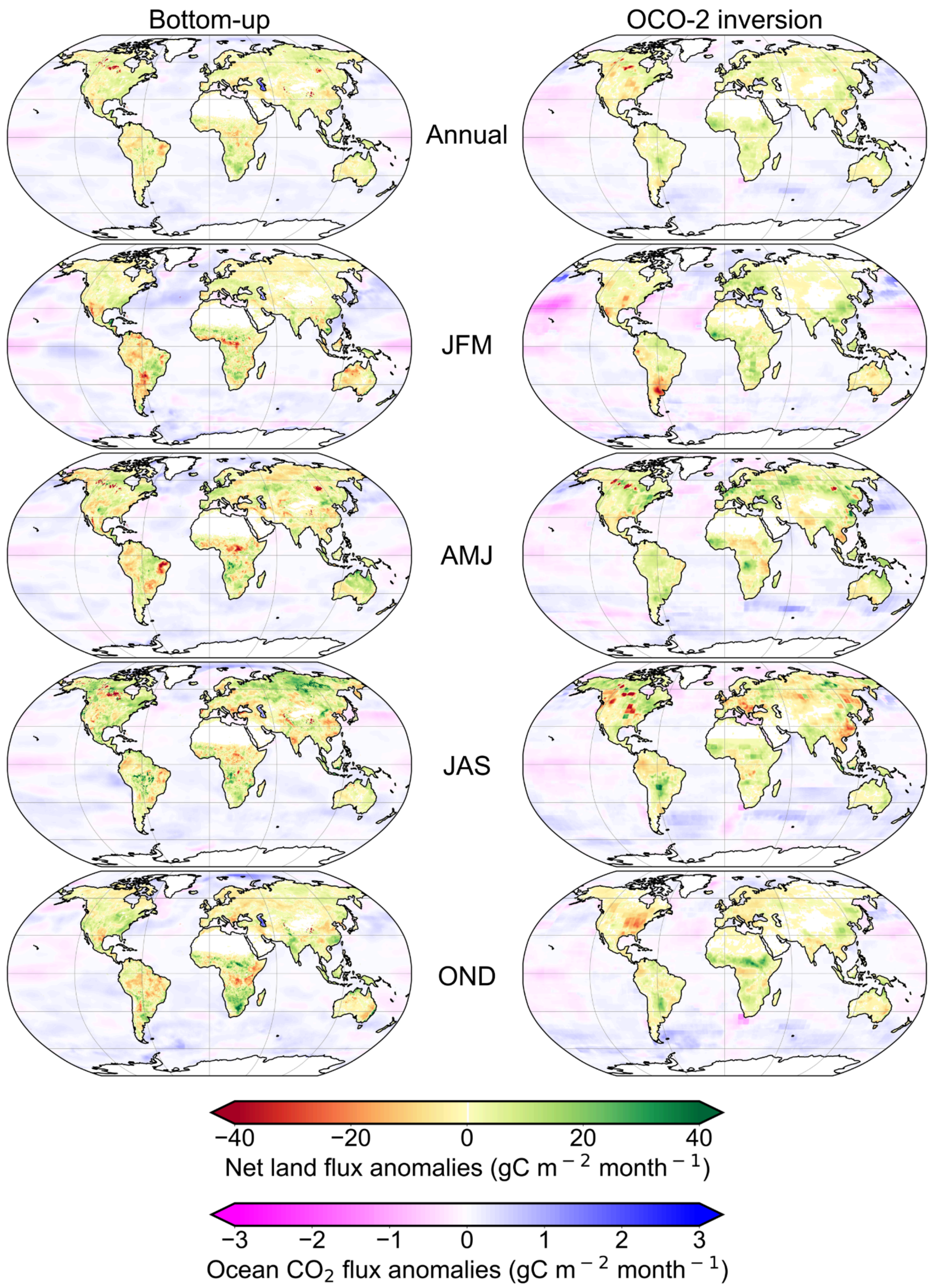

**Fig. S1** **Net land and ocean $CO_2$ flux anomalies for each quarter in 2025 compared with the 2015-2022 average for bottom-up models (left column) and the OCO-2 inversions (right column). Positive values represent increased flux from the atmosphere to the land or ocean (carbon sink).**

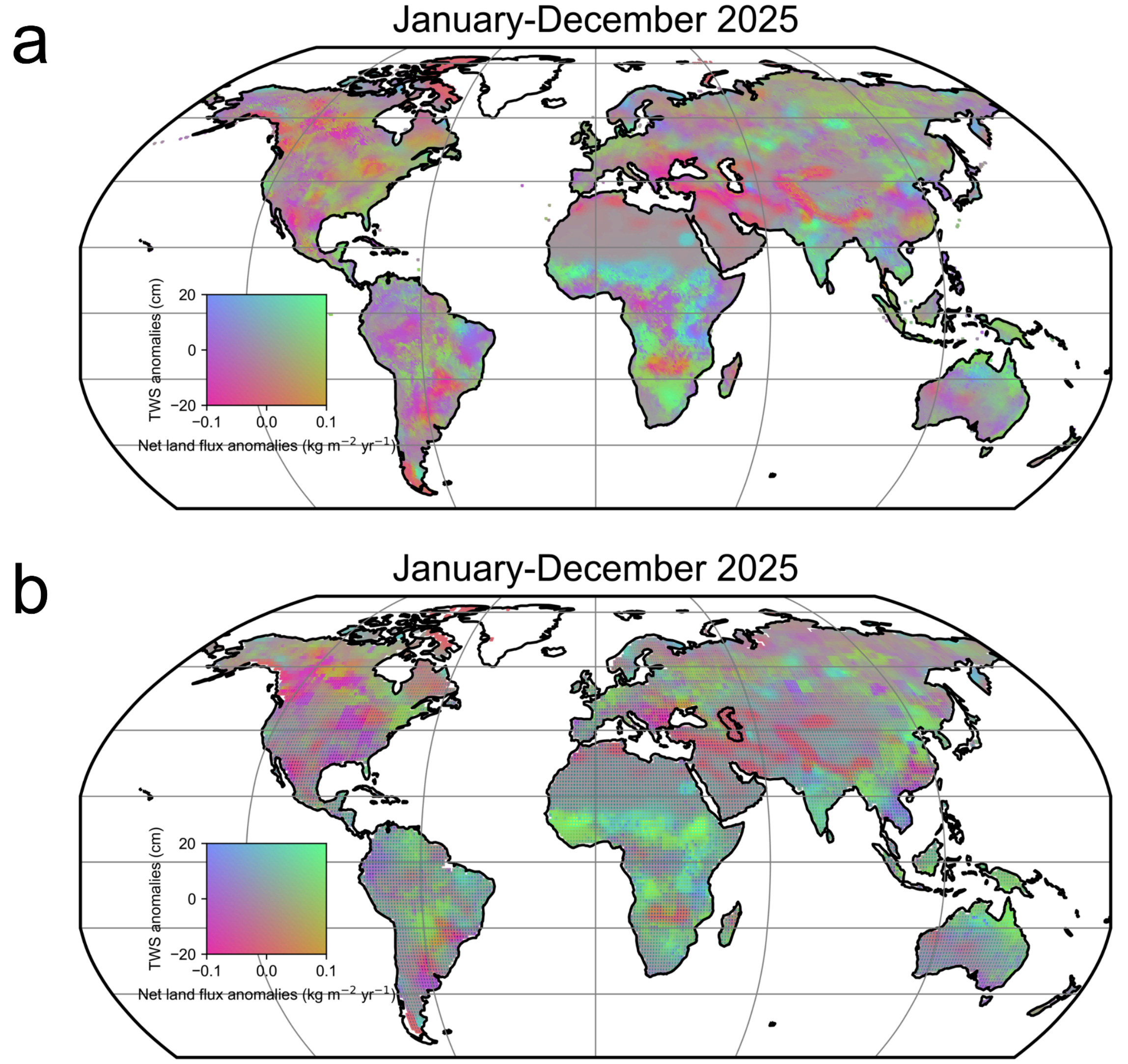

**Fig. S2 Bivariate plots showing co-variations between net land flux anomalies from (a) DGVMs and (b) inversions and total water storage anomalies from the GRACE satellites in 2025.** Green areas show wetter anomalies coincident with more $CO_2$ uptake, and magenta areas show drier anomalies that are coincident with reduced $CO_2$ uptake.